\documentclass[a4paper,UKenglish,cleveref,thm-restate]{lipics-v2021}

\nolinenumbers{}

\newcommand{\parag}[1]{\vspace{-1ex}\subparagraph*{#1}}
\newcommand{\itemname}[1]{\textbf{\textsf{#1}}}

\newcommand{\partfn}{\mathsf{part}}
\newcommand{\bucket}{\mathsf{bucket}}
\newcommand{\slot}{\mathsf{slot}}
\newcommand{\reduce}{\mathsf{reduce}}
\newcommand{\hash}{\mathsf{h}}
\newcommand{\hp}{\mathsf{h}_{\mathsf{p}}}
\newcommand{\Constant}{\mathsf{C}}
\newcommand{\select}{\mathsf{select}}
\newcommand{\free}{F}
\newcommand{\mphf}{\mathsf{H_{mphf}}}
\newcommand{\hi}{\mathsf{hi}}
\newcommand{\lo}{\mathsf{lo}}

\usepackage{ragged2e} % to justify text in abstract
\usepackage{graphicx}
\usepackage{svg}
\usepackage{balance}
\usepackage{xcolor}
\usepackage{xspace}
\usepackage[utf8]{inputenc}
\usepackage{float}
\usepackage{amssymb}
\usepackage{amsmath}
\usepackage{multirow}
\usepackage{booktabs}
\usepackage{makecell}
\usepackage{amsthm}
\usepackage{thmtools}
\usepackage[makeroom]{cancel}
\usepackage{xcolor, soul}
\usepackage{siunitx}
\usepackage{tablefootnote}

\usepackage{fancyvrb}
\VerbatimFootnotes

\usepackage[noEnd=true,indLines=true]{algpseudocodex}
\usepackage[ruled]{algorithm}
\tikzset{algpxIndentLine/.style={draw=black}}

\usepackage{afterpage}

\hypersetup{
    colorlinks=true,
    linkcolor=blue,
    citecolor=blue,
    filecolor=magenta,
    urlcolor=magenta
}

\title{PtrHash: Minimal Perfect Hashing at RAM Throughput}

\author{Ragnar Groot Koerkamp}{ETH Zurich, Zurich, Switzerland}{ragnar.grootkoerkamp@inf.ethz.ch}{https://orcid.org/0000-0002-2091-1237}{ETH Research Grant ETH-1721-1 to Gunnar Rätsch.}


\authorrunning{R. Groot koerkamp}
\Copyright{Ragnar Groot Koerkamp}

\supplementdetails[linktext={https://github.com/RagnarGrootKoerkamp/PtrHash}]{Software}{https://github.com/RagnarGrootKoerkamp/PtrHash}

\acknowledgements{
First, I thank Giulio Ermanno Pibiri for his ongoing feedback in
various stages of this project. Further, I thank Sebastiano Vigna for feedback
from trying to construct PtrHash on \(10^{12}\)~keys and integrating
$\varepsilon$-serde, and lastly I thank Hans-Peter Lehmann and the anonymous
reviewers for feedback on the text.
}


\EventEditors{Petra Mutzel and Nicola Prezza}
\EventNoEds{2}
\EventLongTitle{23rd International Symposium on Experimental Algorithms (SEA 2025)}
\EventShortTitle{SEA 2025}
\EventAcronym{SEA}
\EventYear{2025}
\EventDate{July 22--24, 2025}
\EventLocation{Venice, Italy}
\EventLogo{}
\SeriesVolume{338}
\ArticleNo{19}

\begin{document}

\hypersetup{pageanchor=false}
\maketitle

\begin{abstract}
\label{sec:orgadeced1}
\parag{Motivation.}
Given a set \(K\) of \(n\) keys, a minimal perfect hash function (MPHF) is a
collision-free bijective map \(\mphf\) from \(K\) to \(\{0, \dots, n-1\}\). These
functions have uses in databases, search engines, and are used in bioinformatics
indexing tools such as Pufferfish (using BBHash), and Piscem~(PTHash).
PTHash is also used in SSHash, a data structure on $k$-mers that supports
membership queries. PTHash only takes around $5\%$ of the total
space of SSHash, and thus, trading slightly more space for faster queries is beneficial.
Thus, this work presents a (minimal) perfect hash function that
first prioritizes query throughput, while also allowing efficient construction
for \(10^9\) or more elements using 2.4~bits of memory per key.

\parag{Contributions.}
Both PTHash and PHOBIC first map all $n$ keys to $n/\lambda < n$ \emph{buckets}.
Then, each bucket stores a \emph{pilot} that controls the final hash value of
the keys mapping to it.
PtrHash builds on this by using 1) fixed-width (uncompressed) 8-bit pilots, 2) a construction
algorithm similar to Cuckoo hashing to find suitable pilot values.
Further, it partitions the keys, so that keys in each part map to their own set of \emph{slots}.
PtrHash
3) uses the same number of buckets \emph{and slots} for each part, with 4) a \emph{single}
remap table to map intermediate positions $\geq n$ to $<n$, 5) encoded using per-cacheline Elias-Fano coding.
Lastly, 6) PtrHash supports \emph{streaming} queries, where we use
prefetching to answer a stream of multiple queries more efficiently than one-by-one processing.

\parag{Results.}
With default parameters, PtrHash takse 2.4~bits per key.
On 300~million string keys, PtrHash is as fast or faster
to build than other MPHFs at a similar size, and at least \(2.1\times\) faster to query. When
streaming multiple queries, this improves to \(3.3\times\) speedup over the
fastest alternative, while also being significantly faster to construct.
When using \(10^9\) integer keys instead, query times are as
low as 12~ns/key when iterating in a for loop, or even down to 8~ns/key when using
the streaming approach, just short of the 7.4~ns inverse throughput of random memory accesses.
\end{abstract}

\newpage

\hypersetup{pageanchor=true}
\setcounter{page}{1}

\section{Introduction}
\label{sec:orgebb9721}
Given a set of \(n\) keys \(\{k_0, \dots, k_{n-1}\}\),
a \emph{hash function} maps them to some co-domain \([m] := \{0, \dots, m-1\}\).
When $m\geq n$ and the hash is injective (collision-free), it is also called \emph{perfect}.
When additionally $m=n$ and it is surjective onto \([n]\), it is \emph{minimal}.
Thus, a \emph{minimal perfect hash function}~(MPHF) bijectively maps a set of \(n\) keys onto \([n]\).

\parag{Metrics.}
Various aspects of MPHF data structures can be optimized.
First, one could minimize its space usage and try to
approach the \(\log_2(e)=1.4427\) bits/key lower bound \cite{mehlhorn82_mphf_size}.
Indeed, there are many recent works in this direction, such as Bipartite
ShockHash-RS, which uses under 1.5 bits/key
\cite{shockhash,bipartite-shockhash,phf-thesis},
and \textsc{Consensus}-RecSplit \cite{consensus}, which goes as low as 1.444 bits/key.

In this paper, we focus primarily on optimizing for query throughput and
secondarily on construction speed, while relaxing space usage up to 3 bits/key.
This continues the line of work of FCH \cite{fch}, PTHash \cite{pthash,pthash-2}, and
PHOBIC \cite{phobic}, that all provide relatively fast queries.

\parag{Problem statement.}
Construct a \emph{minimal perfect hash function}
data structure \(\mphf\) that is fast to query, ideally using one memory access
per lookup,
and fast to construct, while staying below 3 bits/key of space.

\parag{Motivation.}
Our main motivating application is to optimize the use of PTHash in SSHash~\cite{sshash}, a data structure to index a set of $k$-mers (sequences
of \(k\) DNA bases).
There, the MPHF only takes around \(5\%\) of the total space. Thus, a slightly
increased space usage of the MPHF has little effect on the total space, while
faster lookups could significantly improve the overall query speed. In this application,
$k$-mers are typically encoded as 64-bit integers, and thus we will focus our
attention on integer keys.

Further applications can be found in domains such as networking \cite{Lu_2006},
databases \cite{Chang_2005}, and
full-text indexing \cite{Belazzougui_2014}, where one could imagine hashing IP addresses,
URLs, or (compact) suffix-trie edge labels.

\parag{Contributions.}
We introduce PtrHash, a minimal perfect hash function that is primarily optimized for
query throughput and construction speed, at the cost of slightly more memory usage.
It builds on the same principles als PTHash(-HEM) and Phobic:
first, keys are partitioned into \emph{parts}. Then, the keys in each part are
further split into \emph{buckets}, and each bucket is assigned a \emph{pilot}
that controls the values (\emph{slots}) that the keys in the bucket hash to.

Compared to PTHash and PHOBIC, the main novelties of PtrHash are:
\begin{enumerate}
\item a fixed number of buckets \emph{and slots} per part, removing the need for
  a part-offset lookup;
\item the use of fixed-width 8-bit \emph{pilots}, so that no compact encoding is needed;
\item a pilot search based on Cuckoo hashing;
\item remapping using a \emph{single} remap table, again simplifying lookups;
\item a remap table based on a per-cacheline Elias-Fano encoding \cite{elias,fano}, CacheLineEF;
\item the use of \emph{prefetching} to \emph{stream} multiple queries in parallel.
\end{enumerate}

\parag{Results.}
When using 300~million string keys, PtrHash with default parameters takes
2.4~bits/key and is nearly as fast to construct as the fastest
other methods, while being much faster to query.
Compared to the next-fastest method to query, PtrHash provides
\(2.1\times\) faster queries when looping naively, or \(3.3\times\) faster when streaming.

When using \(10^9\) integer keys instead, PtrHash can achieve an inverse throughput%
\footnote{For interpretability and consistency with latency numbers, we report
the inverse throughput in nanoseconds per key, rather than keys per second. We
will still refer to this as \emph{throughput}, rather than \emph{inverse
throughput}, following the \href{https://www.intel.com/content/www/us/en/docs/intrinsics-guide/index.html}{intel instruction manual}. }
as low as 12~ns/key when looping over queries, or even 8~ns/key when streaming.

The hardware used for benchmarking has a maximum single-threaded memory bandwidth of 7.4~ns per
cache line. Thus, under the assumption\footnote{This is a strong assumption, and
indeed, PtrHash with the cubic bucket assignment function already slightly breaks this
assumed lower bound.} that almost every query requires
reading at least one new cache line from main memory, our method is close to the
maximum possible query throughput.
Likewise, in a multi-threaded setting, PtrHash can fully saturate the DDR4 memory
bandwidth while answering around 1 query per fetched cache line.

\section{Related work}
\label{sec:orgfe4e2e9}
There is a vast amount of literature on (minimal) perfect hashing, going back to
e.g. \cite{phf-family}. Here we only
give a highlight of recent approaches. We refer the reader to Section 2 of
\cite{pthash-2} and Sections 4 and 8 of the thesis of Hans-Peter Lehmann
\cite{phf-thesis}, which contains a nice overview of different approches
taken by various tools.

\parag{Space lower bound.}
There is a lower bound of \(n \log_2(e)\)~bits to store a minimal perfect hash
function on \(n\) random keys \cite{mehlhorn82_mphf_size}.
To get some feeling for this bound, consider any hash function.
Intuitively the probability that this is
an MPHF is \(n!/n^n\). From this, it follows that at most, around
\(\log_2(n^n/n!)\approx n\log_2(e)\)~bits of information are needed to ``steer'' the hash
function in the right direction.
Now, a naive approach is to use a seeded hash function, and try
\(O(e^n)\) seeds until a perfect hash function is found. However, that is not
feasible in practice.
The method that currently gets closest to the lower bound is
\textsc{Consensus}-RecSplit \cite{consensus}, which goes as low as 1.444 bits/key.

\parag{Bucket placement.}
PtrHash builds on methods that first group the keys into
buckets of a few keys. Then, keys in the buckets are assigned their hash value
one bucket at a time, such that newly assigned values do not collide with
previously taken values. All methods iterate different possible key assignments
for each bucket until a collision-free one is found, but differ in the way
hash values are determined. To speed up this search,
large buckets are assigned a hash before small buckets, since smaller buckets
are easier to place when many slots are already taken.

FCH \cite{fch} uses a fixed number of bits to encode the seed for each bucket and
uses a \emph{skew} distribution of bucket sizes. The seed stored in each bucket
determines how far the keys are \emph{displaced} (rotated) to the right from their
initially hashed positions. A fallback hash can be used if needed, and
construction can fail if that also does not work. CHD \cite{chd} uses uniform
bucket sizes, but uses a variable-width encoding for the seeds.
PTHash \cite{pthash} combines these two ideas and introduces a number of
compression schemes for the seed values, that are called \emph{pilots}. Instead of
directly generating an MPHF, it first generates a PHF to \([n']\) for
\(n'=n/\alpha \approx n/0.99\), and values mapping to positions \(\geq n\) are \emph{remapped} to
the skipped values in \([n]\). PTHash-HEM \cite{pthash-2} first partitions the keys, and uses this
to build multiple parts in parallel. This also enables external-memory construction.
Lastly, PHOBIC \cite{phobic} improves from the simple \emph{skew} distribution of
FCH to an \emph{optimal bucket assignment function}, which speeds up construction and
enables smaller space usage. Secondly, it partitions the input into parts of
expected size
2500 and uses the same number of buckets for each part. Then, it uses that the
pilot values of the \(i\)'th bucket of each part follow the same distribution, and
encodes them together. Together, this saves 0.17 bits/key over PTHash.
Lastly, some of the ideas in PtrHash (fixed 8-bit pilots and cuckoo hashing)
have been independently proposed in \cite{phobic-thesis}.

\section{PtrHash}
\label{sec:orgce4a522}

The core design goal of PtrHash\footnote{The
PT in PTHash stand for \emph{Pilot Table}. The
author of the present paper mistakenly understood it to stand for Pibiri and
Trani, the authors of the PTHash paper. Due to the current author's
unconventional last name, and PTGK not sounding great, the first initial (R) was
appended instead, doubling as a hint that PtrHash is written in Rust. As things go, nothing is as permanent as a temporary name.
Furthermore, we follow the Google style guide and avoid a long run of uppercase
letters, and write PtrHash instead of PTRHash.}
is to simplify PTHash to speed up both query speed
and construction time, at the cost of possibly using slightly more memory.

\subsection{Overview}
\label{sec:org06ce748}

\begin{figure}[t]
\centering
\includegraphics[width=\linewidth]{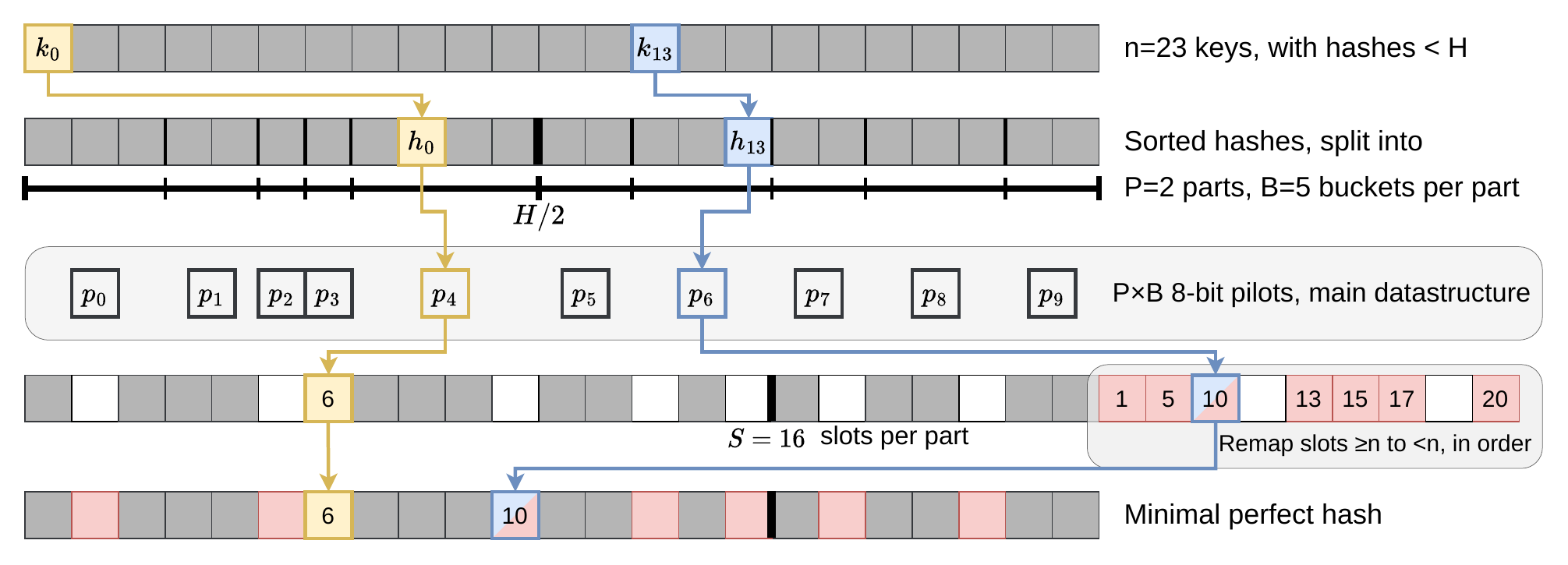}
\caption{\label{overview}Overview of PtrHash on \(n=23\) keys. The keys are
  hashed into \([H] = [2^{64}]\) and this range is split into \(P=2\) parts and
  \(B=5\) buckets per part. The key highlighted in yellow has a the 9'th
  smallest hash, and ends up in \emph{bucket} 4 (starting at index 0). The corresponding
  \emph{pilot} $p_4$ hashes the key to \emph{slot} 6.
  The array of pilots (grey background) is the main component of the PtrHash
  data structure, and ensures that all keys hash to different slots.
  The blue key has a hash in the second
  part (upper half) of hashes, in bucket 6. It gets hashed to slot 25, which is
  larger than the number of keys $n=23$. Thus, it is \emph{remapped} (along with
  the other red cells) into an
  empty slot $<n$ via a (compressed) list of free slots, which is the second
  main component of the data structure.}
\vspace{-1em}
\end{figure}

Before going into details, we first briefly explain the fully constructed
PtrHash data structure and how to query it, see \cref{overview}. We also
highlight differences to PTHash \cite{pthash} and PHOBIC \cite{phobic}.

\parag{Parts and buckets.}
The input is a set of \(n\) \emph{keys} \(\{k_0, \dots, k_{n-1}\}\) that we want to hash to
\(n\) \emph{slots} \([n]:=\{0, \dots, n-1\}\).
We first hash the keys using a 64-bit hash function \(\hash\) into
\(\{\hash(k_0), \dots, \hash(k_{n-1})\}\). The total space of hashes \([2^{64}]\)
is equally partitioned into \(P\) \emph{parts}, and the part of a key is easily found
as \(\left\lfloor P\cdot \hash(k_i) / 2^{64}\right\rfloor = \hi(P\cdot \hash(k_i)) \)
\cite{fast-range}, where $\hi(a\cdot b)$ returns the high 64 bits of the product
of two 64-bit integers, and likewise, $\lo(a\cdot b)$ returns the low 64 bits.
Then, the expected \(n/P\) keys in each part are further split into exactly \(B\) non-uniform \emph{buckets}:
each key has a \emph{relative position} \(x\) inside the part, and this is passed through
a \emph{bucket
assignment function} \(\gamma: [0,1)\mapsto[0,1)\) such as \(\gamma(x)= x^2\)
that controls the distribution of expected bucket
sizes \cite{phobic}, as explained in detail in \cref{sec:bucket-fn}.
The result is then scaled to a bucket index in \([B]\):
\begin{align}
\begin{split}
  \partfn(k_i) &:=  \hi(P\cdot \hash(k_i)),\\
  x &:= \lo(P\cdot \hash(k_i))/2^{64},\\
  \bucket(k_i) &:= \hi(B\cdot (2^{64}\cdot\gamma(x))).
\end{split}\label{eq:partbucket}
\end{align}

\parag{Slots and pilots.}
Now, the goal and core of the data structure is to map the \(n/P\) expected keys in each part to \(S\approx
(n/P)/\alpha\) \emph{slots}, where \(\alpha\approx 0.99\) gives us \(\approx
1\%\) extra slots to play with. The pilot for each bucket controls to which slots its keys map.
PtrHash uses fixed-width 8-bit \emph{pilots} \cite{phobic-thesis} \(\{p_0, \dots,
p_{P\cdot B-1}\}\), one for each bucket. Specifically, key \(k_i\) in bucket \(\bucket(k_i)\) with pilot \(p_{\bucket(k_i)}\)
maps to slot
\begin{equation}
  \slot(k_i) := \partfn(k_i) \cdot S + \reduce(\hash(k_i) \oplus \hp(p_{\bucket(k_i)}), S),\label{eq:slot}
\end{equation}
where \(\reduce(\cdot, S)\) maps the random 64-bit integer into \([S]\) as
explained below, and $\oplus$ denotes xor.

Compared to PHOBIC and PTHash(-HEM) \cite{pthash-2}, there are two differences
here.
First, while we still split the input into parts, we assign each part not
only the same number of buckets, but also the
\emph{same} number of slots, instead of scaling the number of slots with the
\emph{actual} size of each part. This removes the need store a prefix sum
of part sizes, and avoids one memory access at query time to look up the
offset of the key's part.
This idea was recently independently introduced as \emph{$\varepsilon$-cost
sharding}~\cite{eps-cost-sharding}.
Second, previous methods search for arbitrary large
pilot values that require some form of compression to store efficiently. Our
8-bit pilots can simply be stored in an array so that lookups are simple.

We now go over some specific details.

\parag{Hash functions.}
The 8-bit pilots \(p_b\) are hashed into pseudo-random 64-bit integers by
using FxHash~\cite{fxhash} for \(\hp\),
which simply multiplies the pilot with a \emph{mixing constant} \(\Constant\) after
xoring by a global seed:
\begin{equation}
\hp(p) := \Constant \cdot (p \oplus \mathrm{seed}).
\end{equation}

When the keys are 64-bit integers, we use this same FxHash algorithm to hash
them (\(\hash(k) := \Constant\cdot k\)), since multiplication by an odd constant is
invertible modulo \(2^{64}\) (since $\gcd(C, 2^{64})=1$) and
hence collision-free.
For other types of keys, the hash function depends on the number of elements. When the
number of elements is not too far above \(10^9\), the probability of hash
collisions with a 64-bit hash function is sufficiently small, and, following
PHast \cite{phast}, we use
the 64-bit variant of GxHash \cite{gxhash}, a hash function based on AES
hardware instructions.
When the number of keys goes beyond \(2^{32} \approx 4\cdot 10^9\), the
probability of 64-bit hash collisions increases. In this case, we use the
128-bit variant of GxHash.
The high 64-bits determine the part and bucket in \cref{eq:partbucket}, and the low
64-bits are used in \cref{eq:slot} to determine the slot.

\parag{The reduce function.}
To obtain the slot inside the current part, we must reduce the hash based on the
key and its pilot to a number in $\{0, \dots, S-1\}$.
One way of doing this is to use ``fast mod''~\cite{fast-mod}, which uses two multiplications
when the modulus (the number of slots per part $S$) is less than $2^{32}$.

When $S$ is a power of two, we can instead use $\reduce(x, S) = \hi(C\cdot x)
\bmod S$, which only needs a single multiplication and a bitmask.
The multiplication by the mixing constant $C$ ensures that all bits of $x$ are used.
In practice, this is the method we use.

When the number of parts is small, a drawback of
limiting $S$ to powers of two is that
this could cause up to $50\%$ empty slots.
In this case, fast mod can be used for reliability.
Then,
that $S$ must \emph{not} a power of two, so that $x\bmod S$ depends on all%
\footnote{Only depending on the $\lg_2 S$ low bits is
not good enough, since the $\partfn$ and $\bucket$ functions only depend on the
high $\lg_2(P\cdot B)$ bits, leaving some bits in the middle usused.}
bits of $x$. Additionally, we can only use a single part, simplifying queries.

\parag{Remapping.} Since each part has slightly (\(\approx 1\%\)) more slots than keys, some keys will map to an
index \(\geq n\), leading to a \emph{non-minimal} perfect hash function. To fix this,
those are \emph{remapped} back into the ``gaps'' left behind in slots \(<n\) using a
(possibly compressed) lookup table. This is explained in detail in \cref{remapping}.

Whereas PTHash-HEM uses a separate remap \emph{per part}, PtrHash only has a single
``global'' remap table.

\parag{Construction.} The main difficulty of PtrHash is during construction (\cref{sec:construction}), where we must find values of the
pilots \(p_j\) such that all keys indeed map to different slots.
Like other methods, PtrHash processes multiple parts in parallel.
Within each part, it sorts the buckets from large to
small and ``greedily'' assigns them the smallest pilot value that maps the keys in
the bucket to slots that are still free.
Unlike other methods though, PtrHash only allows pilots up to \(255\). When no
suitable pilot is found, we use a method similar to (blocked) Cuckoo hashing
\cite{cuckoo-hashing,dary-cuckoo-hashing}: a pilot with a minimal number of collisions is chosen,
and the colliding buckets are ``evicted'' and will have to search for a new pilot.

\parag{Parameter values.}
In practice, we usually use \(\alpha=0.99\).
Similar to PHOBIC, the number of buckets per part is set to \(B = \lceil(\alpha\cdot
S)/\lambda\rceil\), where \(\lambda\) is the expected size of each bucket and is around
\(3\) to \(4\).
The number of parts is \(P=\lceil n/(\alpha S)\rceil\).
Smaller parts fit better in cache and hence are faster to construct, while too
small parts have too much variance in their size, causing some parts to possibly have
more than \(S\) keys in them. Thus, we would like to choose \(S\) as the smallest size for which
the probability that any part is over-subscribed is sufficiently small.
Vigna \cite[eq. 3]{eps-cost-sharding} shows that in practice, the following
formula works well:
\begin{equation*}
  P \approx n/(\alpha S) \leq \frac{n\varepsilon^2/2}{\ln\left(n\varepsilon^2/2\right)},
\end{equation*}
where we use $\varepsilon = (1-\alpha)/2$ to ensure that all parts have at last
half of the average number of free slots.
For $\alpha=0.99$, this reduces to
\begin{equation}
  \alpha S \geq 80\,000 \cdot \ln(n/80\,000),
\end{equation}
and so this is the number of key per part $\alpha S$ we choose, with a minimum of
$80\,000$ for when $n\leq 80\,000$.

\parag{Streaming queries.} PtrHash supports \emph{streaming} queries, where multiple
queries are processed in parallel. This allows prefetching pilots from
memory, and thus increases throughput and better uses the available memory bandwidth.
This is explained and evaluated in \cref{sec:throughput}.

\parag{Sharding.} When the number of keys is so large that their hashes do not
fit into memory, one of three sharding strategies can be used: in-memory,
on-disk, or hybrid. These are explained and evaluated in \cref{sec:sharding}.

\subsection{Construction}
\label{sec:construction}
Both PTHash-HEM and PHOBIC first partition the keys into parts, and then build
an MPHF part-by-part, optionally in parallel on multiple threads.
Within each part, the keys are randomly split into
\emph{buckets} of average size \(\lambda\) (\cref{overview}).
Since $\lambda \leq 4$ in practice, the variance on bucket sizes is quite large.
Thus, the buckets are sorted from large to small, and one-by-one \emph{greedily} assigned a
\emph{pilot}, such that the keys in the bucket map to \emph{slots} not yet covered by earlier buckets.

As more buckets are placed, there are fewer remaining empty slots, and searching for pilots becomes harder.
Hence, PTHash uses \(n/\alpha > n\) slots
to ensure there sufficiently many empty slots for the last pilots. This speeds
up the search and reduces the values of the pilots.
PHOBIC, on the other hand, uses relatively small parts of expected size 2500, so that
the search for the last empty slot usually should not take much more than 2500 attempts.
Nevertheless, a drawback of the greedy approach is that pilots values have an uneven
distribution, making it somewhat harder to compress them while still allowing
fast access (e.g., requiring the interleaved coding of PHOBIC).

\parag{Hash-evict\footnote{We would have preferred to call this method hash-displace, as
\emph{displace} is the term used instead of \emph{evict} in e.g. the Cuckoo filter~\cite{cuckoo-filter}.
Unfortunately, \emph{hash and displace} is already taken by compressed
hash-and-displace~\cite{hash-displace,chd}}.} In PtrHash, we instead use \emph{fixed width}, single byte pilots. To achieve
this, we use a technique resembling Cuckoo hashing \cite{cuckoo-hashing} that
was also independently found in \cite[Section 4.5]{phobic-thesis}.
As before, buckets are greedily \emph{inserted} from large to small. For some buckets,
there may be no pilot in \([2^8]\) such that all its keys map to empty slots. When
this happens, a pilot is found with the lowest weighted number of \emph{collisions}.
The weight of a collision with an element of a bucket of size \(s\) is \(s^2\), to prevent
\emph{evicting}%
\footnote{We would have preferred to call this method hash-displace, as
\emph{displace} is the term used instead of \emph{evict} in e.g. the cuckoo filter~\cite{cuckoo-filter}.
Unfortunately, \emph{hash and displace} is already taken
by hash-and-displace~\cite{hash-displace,chd}.}
large buckets, as those are harder to place.
The colliding buckets are evicted by emptying the slots they map to and
pushing them back onto the priority queue of remaining buckets.
Then, the new bucket is inserted, and the next largest remaining or evicted
bucket is processed.

In order to efficiently search for pilot vectors, we use a bitvector of taken slots.
Additionally, we avoid infinite loops of evicted buckets by storing the 16 most
recently placed buckets, and never displacing those.

\subsection{Bucket Assignment Functions}
\label{sec:bucket-fn}

\begin{figure}[t]
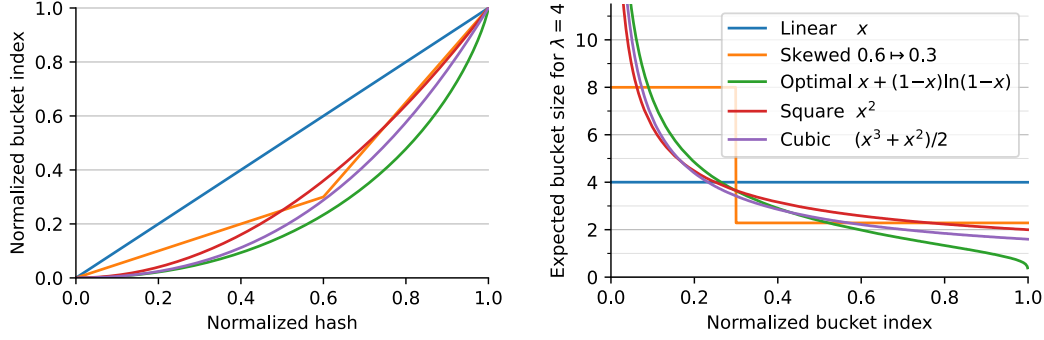

  \centering
  \begin{subfigure}[t]{0.49\linewidth}
    \includesvg[width=1\linewidth]{plots/bucket-fn}
  \end{subfigure}
  \hfill
  \begin{subfigure}[t]{0.49\linewidth}
    \includesvg[width=1\linewidth]{plots/bucket-size}
  \end{subfigure}
  \caption{\label{bucket-fn}The left shows various bucket assignment functions \(\gamma\), such as the piecewise linear function (skewed) used by FCH and PTHash, and the optimal function introduced by PHOBIC. Flatter slopes at \(x=0\) create larger buckets, while steeper slopes at \(x=1\) create more small buckets, as shown on the right, as the distribution of expected bucket sizes given by \((\gamma^{-1})'\) when the expected bucket size is \(\lambda=4\).}
\end{figure}

During construction, slots fill up as more buckets are
placed. Because of this, the first buckets are much easier to place than the
later ones, when only few empty slots are left.
To compensate for this, we can introduce an uneven distribution of bucket
sizes, so that the first buckets are much larger and the last buckets
are smaller.
FCH \cite{fch} accomplishes this by a \emph{skew} mapping that assigns \(60\%\) of the
elements to \(30\%\) of the
buckets, so that those \(30\%\) are \emph{large} buckets while the remaining \(70\%\)
is \emph{small} (\cref{bucket-fn}). This is also the scheme used by PTHash.

\parag{The optimal bucket function.}
PHOBIC \cite{phobic} provides a more thorough analysis and uses the optimal function
\(\gamma_p(x) = x + (1-x)\ln (1-x)\) when the target load factor is \(\alpha=1\).
A small modification is optimal for \(\alpha<1\) \cite[Appendix B]{phobic-full},
but for simplicity we only consider the original \(\gamma_p\).
This function has derivative 0 at \(x=0\), so
that many \(x\) values map close to 0.
In practice, this causes the largest buckets to have size much larger than \(\sqrt S\).
Such buckets are hard to place, because by the birthday paradox they are likely
to have multiple elements hashing to the same slot. To fix this, PHOBIC ensures the
slope of \(\gamma\) is at least \(\varepsilon=1/\big(5 \sqrt S\big)\) by using
\(\gamma_{p,\varepsilon(x)} = x + (1-\varepsilon)(1-x)\ln(1-x)\) instead.
For simplicity in the implementation, we fix \(\varepsilon = 1/{2^8}\), which
works well in practice.

\parag{Approximations.}
For PtrHash, we aim for high query throughput, and thus we would like to only
use simple computations and avoid additional lookups as
much as possible.
To this end, we replace the \(\ln (1-x)\) by its
first order Taylor approximation at \(x=0\), \(\ln(1-x) \approx -x\), giving
the quadratic \(\gamma_2(x) := x^2\). Using the second order approximation \(\ln(1-x) \approx
-x-x^2/2\) results in the cubic \(\gamma(x) = (x^2+x^3)/2\). This version again
suffers from too large buckets, so in practice we use \(\gamma_3(x) =
\frac{2^8-1}{2^8}\cdot (x^2+x^3)/2 + \frac{1}{2^8}\cdot x\).
We also test the trivial $\gamma_1(x):=x$.

These values can all be computed efficiently by using that the input and output
of \(\gamma\) are 64-bit unsigned integers representing a fraction of \(2^{64}\),
so that e.g. \(x^2\) can simply be computed as $\hi(x\cdot x)$.

\subsection{Remapping using CacheLineEF}
\label{remapping}
Like PTHash, PtrHash uses a parameter \(0<\alpha\leq 1\) to use a total of
\(n'=n/\alpha\) slots, introducing \(n'-n\) additional free slots.
As a result of the additional slots, some, say \(R\), of the keys will map to positions \(n\leq
q_0<\dots< q_{R-1}< n'\), causing the perfect hash function to not be \emph{minimal}.

\parag{Remapping.} Since there are a total of \(n\) keys, this means there are exactly \(R\) empty
slots~(``gaps'') left behind in \([n]\), say at positions \(L_0\) to \(L_{R-1}\).
We \emph{remap} the keys that map to positions \(\geq n\) to the empty slots at
positions \(< n\) to obtain a \emph{minimal} perfect hash function.

A simple way to store the remap is as a plain array \(\free\), such that
\(\free[q_i-n] = L_i\).
PTHash encodes this array using Elias-Fano coding \cite{elias,fano}, after setting undefined
positions of \(\free\) equal to their predecessor.
The benefit of a plain \(\free\) array is fast and cache-local lookups, whereas
Elias-Fano coding provides a more compact encoding that typically requires multiple
lookups to memory.

\begin{figure}[t]
\centering
\includegraphics[width=\linewidth]{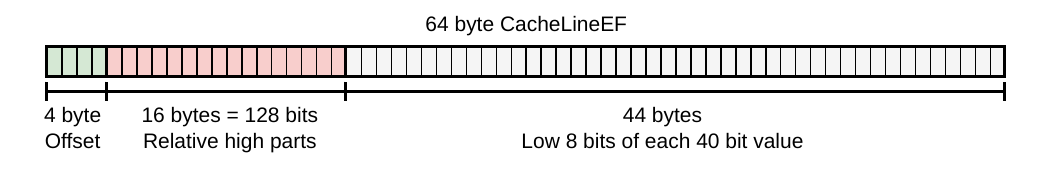}
\caption{\label{cacheline-ef}Overview of the CacheLineEF data structure.}
\end{figure}

\parag{CacheLineEF.}
We would like to answer each query by reading only a single cache line from
memory. To do this, we use a method based on \emph{interleaving} data.
First, the list of non-decreasing \(\free\) positions is split into chunks of
\(C=44\) values \(\{v_0, \dots, v_{43}\}\), with the last chunk possibly containing fewer values.
We assume that values are at most 40~bits, and that the average difference
between adjacent values in each chunk is not more than 500.
Then, each chunk is encoded into 64 bytes that can be stored as single cache
line, as shown in \cref{cacheline-ef}.

We first split all values into their 8 \emph{low} bits (\(v_i \bmod 2^8\)) and 32
\emph{high} bits (\(\lfloor v_i/2^8\rfloor\)). Further, the high part is split into an
\emph{offset} (the high part of \(v_0\)) and the \emph{relative} high part:
\begin{equation}
v_i =
2^8\cdot\underbrace{\lfloor v_0/2^8\rfloor}_{\text{Offset}} +
2^8\cdot \underbrace{\left(\lfloor v_i/2^8\rfloor - \lfloor
v_0/2^8\rfloor\right)}_{\text{Relative high part}}
+\underbrace{(v_i\bmod 2^8)}_{\text{Low bits}}.
\label{eq:clef}
\end{equation}
This is stored as follows.
\begin{itemize}
\item First, the 32 bit offset \(\lfloor v_0/2^8\rfloor\) is stored.
\item Then, the relative high parts are encoded into \(128\) bits. For each \(i\in[44]\), bit \(i + \lfloor
  v_i/2^8\rfloor - \lfloor v_0/2^8\rfloor\) is set to 1.
Since the \(v_i\) are increasing, each \(i\) sets a distinct bit, for a total of 44~set bits.
\item Lastly, the low 8~bits of each \(v_i\) are directly written to the 44~trailing bytes.
\end{itemize}

\parag{Lookup.} The value at position \(i\) is found by summing the terms of
\cref{eq:clef}. The offset and low bits can be read directly.
This relative high part can be found as \(2^8\cdot(\select(i)-i)\), where \(\select(i)\) gives
the position of the \(i\)'th 1 bit in the 128-bit-encode relative high parts. In practice, this can be implemented
efficiently using a \texttt{popcount} to go into the high or low half, followed by the \texttt{PDEP} instruction\footnote{Unfortunately, while AMD Zen
2 does support this instruction, it is very slow in practice.} provided by the BMI2 bit manipulation
instruction set~\cite{fast-select}.

\parag{Limitations.} CacheLineEF uses \(64/44\cdot 8 = 11.6\)~bits per value, which is
more than the usual Elias-Fano, which for example takes \(8+2=10\)~bits per value for data
with an average \emph{stride} (gap between consecutive integers) of \(2^8\).
Furthermore, values are limited to 40~bits, covering \(10^{12}\) items.
The range could be increased to 48~bit numbers by storing 5 bytes of the
offset, but this has not been necessary so far.
Lastly, each CacheLineEF can only span a range of around \((128-44)\cdot 2^8 =
21\,504\), or an average stride of 500.
This means that for PtrHash, we only use CacheLineEF when \(\alpha\leq 0.99\), so that the
average distance between empty slots is 100 and the average stride of 500 is
not exceeded in practice. When \(\alpha > 0.99\), a simple plain array can be used
without much overhead.

\section{Results}
\label{sec:orgbf28892}
We now evaluate PtrHash construction and query throughput for
different parameters, and compare PtrHash to other minimal perfect hash functions.
All experiments are run on an Intel Core i7-10750H CPU with 6 cores and
hyper-threading disabled.
The frequency is pinned to 2.6~GHz.
Cache sizes are 32~KiB L1 and 256 KiB L2 per core, and 12~MiB shared L3 cache. Main
memory is 64~GiB DDR4 at 3200~MHz, split over two 32~GiB banks.

In \cref{construction-eval}, we compare the effect of various parameters and configurations on the
size, construction speed, and query speed of PtrHash.
In \cref{sec:comparison}, we compare PtrHash to other methods.

Further, \cref{throughput-evaluation} evaluates the effect of prefetching with
batching and streaming queries. We select streaming with prefetching 32
iterations ahead as the default. We also show that in a multi-threaded setting,
this can fully exhaust the available memory bandwidth.

Lastly, in \cref{sec:sharding-eval} we state the results of constructing PtrHash
on 50 billion keys using various sharding strategies.

\subsection{Construction}
\label{construction-eval}

The construction experiments use \(10^9\) random 64-bit integer keys,
for which the data structure takes
around 300 MB and thus is much larger than L3 cache. Unless otherwise mentioned,
construction is in parallel using 6 cores.
For the query throughput experiments, we also test on
20 million keys, for which the data structure take around
6 MB and easily fit in L3 cache.
To avoid the time needed for hashing keys, and since our motivating application
is indexing $k$-mers that fit in 64~bits, we always use random 64-bit integer keys, and hash them using FxHash.

Without using the external memory construction,
memory usage during construction is dominated by the size of the input keys and
their hashes, which are typically much larger than the few bits per key needed
for the construction itself.

\subsubsection{Bucket Functions}
\label{sec:orge11d60c}

\begin{figure}[t]
  \centering
\includesvg[width=\linewidth]{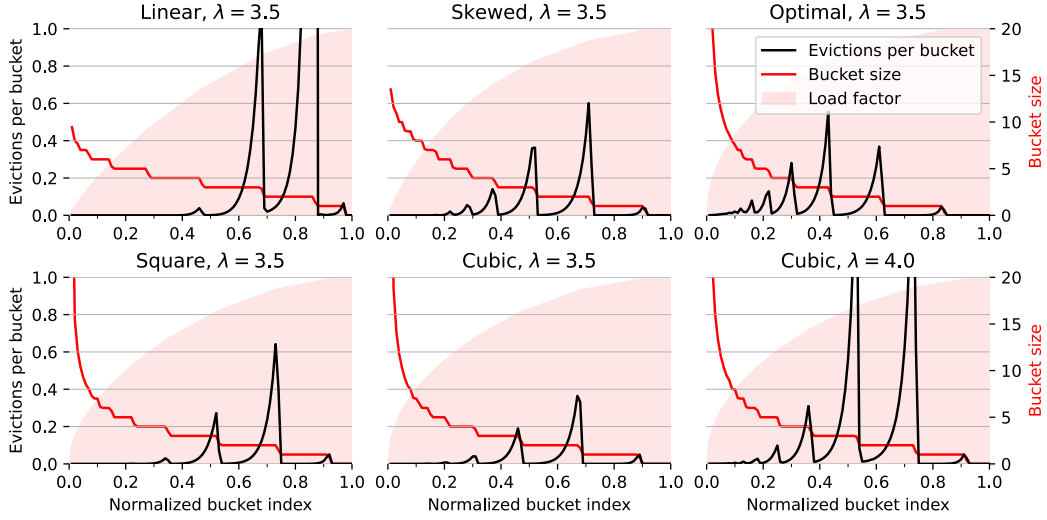}
\caption{\label{bucket-fn-plot}Bucket size distribution (red) and average number of evictions (black) per additionally placed bucket during construction of the pilot table, for different bucket assignment functions. Parameters are \(n=10^9\)~keys, \(S=2^{18}\) slots per part, and \(\alpha=0.99\), and the red shaded load factor ranges from 0 to \(\alpha\). In the first five plots \(\lambda=3.5\) so that the pilots take 2.29~bits/key. For \(\lambda=4.0\) (bottom-right), the linear, skewed, and optimal bucket assignment functions cause endless evictions, and construction fails. The cubic function does work, resulting in 2.0~bits/key for the pilots.}
\vspace{-1em}
\end{figure}

In \cref{bucket-fn-plot}, we compare the performance of different bucket assignment
functions \(\gamma\) in terms of the bucket size distribution and the number of
evictions for each additionally placed bucket.
We see that the linear \(\gamma_1(x) = x\) has a lot of evictions for the last
buckets of size \(3\) and \(2\), but like all methods it is fast for the last
buckets of size \(1\) due to the load factor \(\alpha < 1\). The optimal
distribution of PHOBIC performs only slightly better than the skewed one of FCH and
PTHash, and can be seen to create more large buckets since the load factor
increases fast for the first buckets.
The cubic~\(\gamma_3\) is clearly much better than all other functions, and is
also tested with larger buckets of average size \(\lambda = 4\), where all other
functions fail.

In the remainder, we will test the linear \(\gamma_1\) for simplicity and lookup
speed, and the cubic \(\gamma_3\) for space efficiency.
\subsubsection{Tuning Parameters for Construction}
\label{sec:org9f908d8}

\begin{figure}[t]
\centering
\includesvg[width=\linewidth]{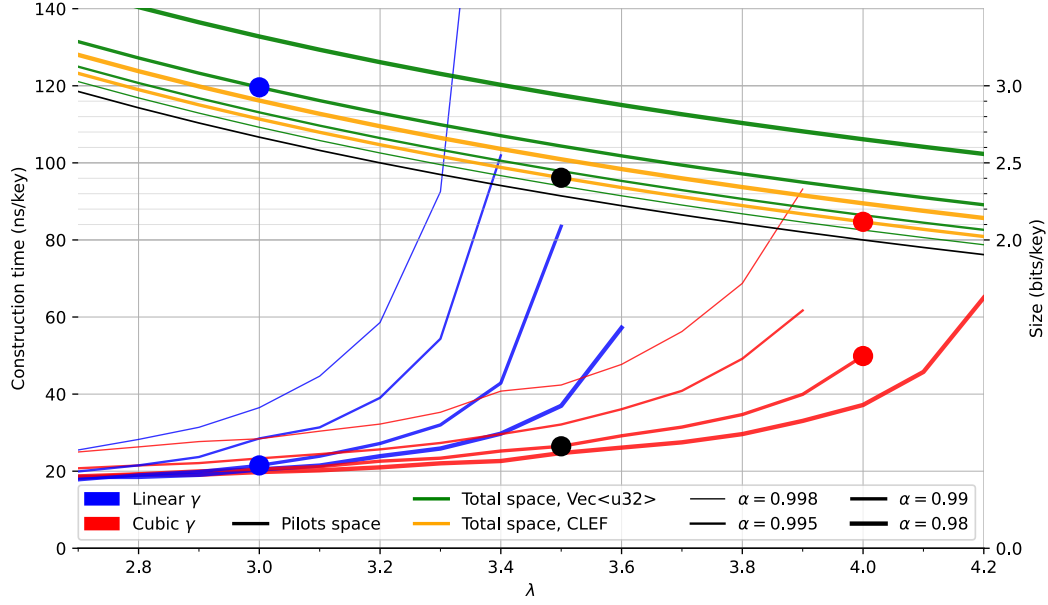}
\caption{\label{fig:construction}This plot shows the construction time (blue and
  red, left axis) and data structure size (black, green, and yellow, right axis)
  as a function of \(\lambda\) for \(n=10^9\) keys. Parallel construction time
  on 6 threads is shown for both the linear and cubic \(\gamma\), and for
  various values of \(\alpha\) (thickness). The curves stop because construction
  times out when \(\lambda\) is too large. For each \(\lambda\), the black line
  shows the space taken by the array of pilots. For larger \(\lambda\) there are
  fewer buckets, and hence the pilots take less space. The total size including
  the remap table is shown in green (plain vector) and yellow (CacheLineEF) for
  various \(\alpha\). The blue (fast), black (default), and red (compact) dots highlight the chosen parameter configurations.}
\end{figure}

In \cref{fig:construction} we compare the multi-threaded construction time and space usage of PtrHash on
\(n=10^9\) keys for
various parameters \(\gamma\in \{\gamma_1, \gamma_3\}\), \(2.7\leq \lambda\leq 4.2\),
\(\alpha\in \{0.98, 0.99, 0.995, 0.998\}\), and plain remapping or CacheLineEF.
We see that for fixed \(\gamma\) and \(\alpha\), the construction time appears to
increase exponentially as \(\lambda\) increases. At too large $\lambda$, some
parts fail to build after a total of $10S$ evictions, which is a hard limit we
impose to avoid running into eviction cycles.
Load factors \(\alpha\) closer to \(1\) (thinner lines) achieve smaller overall data
structure size, but take longer to construct and time out at smaller \(\lambda\).
The cubic \(\gamma_3\) is faster to construct than the identity \(\gamma_1\) for
small \(\lambda \leq 3.5\). Unlike \(\gamma_1\), it also scales to much larger
\(\lambda\) up to \(4\), and thereby achieves significantly smaller overall size.

We note that for small \(\lambda\), construction time does converge to around 19!ns/key.
A rough time breakdown is that for each key, 1~ns is spent on hashing, 5~ns
on sorting all the keys, 12~ns to search for pilots, and lastly 1~ns on remapping
to empty slots.

\parag{Recommended parameters.}
Based on these results, we choose three sets of parameters for further
evaluation, as indicated with blue, black, and red dots in \cref{fig:construction}:
\begin{itemize}
\item \itemname{Fast} (blue), aiming for query speed: using the linear \(\gamma_1\), \(\lambda=3.0\), \(\alpha=0.99\), and a plain
vector for remapping.
Construction takes only just over 20~ns/key, close to the apparent lower
bound, and space usage is 3 bits/key. This can be used when \(n\) is small, or
more generally when memory usage is not a bottleneck.
\item \itemname{Default} (black), a trade-off between fast construction and small
space: using cubic \(\gamma_3\), \(\lambda=3.5\), and \(\alpha=0.99\), with
CacheLineEF remapping.
\item \itemname{Compact} (red), aiming for small space: using the cubic \(\gamma_3\), \(\lambda=4.0\), \(\alpha=0.99\), and
CacheLineEF remapping. Construction now takes around 50~ns/key, but the data
structure only uses 2.12 bits/key. In practice, this configuration sometimes
ends up in endless eviction cycles, and $\lambda=3.9$ may be better.
\end{itemize}
\subsubsection{Remap}
\label{sec:orgece074a}
\begin{table}[t]
\caption{\label{tab:remap}Comparison of space usage (bits/key) and query throughput
  (ns/query) of PtrHash when using the recommended parameters with different
  remap structures. Query throughput is shown both for perfect hashing (without
  remap), and for minimal perfect hashing (with remap). Additionally, query
  throughput is shown both for a for-loop and for streaming.}
\centering
\setlength{\tabcolsep}{5pt}
\begin{tabular}{lrrrlrrr}
\toprule
Configuration & Pilots & \multicolumn{2}{c}{Query PHF} & \multicolumn{2}{c}{Remap} &
\multicolumn{2}{c}{Query MPHF} \\
\cmidrule(lr){3-4}
\cmidrule(lr){5-6}
\cmidrule(lr){7-8}
 & Space & Loop & Stream & Type & \hspace{-4em} Space& Loop & Stream\\
\midrule
& 2.67 & 11.5 & 8.6 & Vec<u32> & 0.33 & 12.5 & 8.8\\
Fast &&&& CacheLineEF          & 0.12 & 12.9 & 8.8\\
 \multicolumn{4}{l}{\(\alpha=0.99\), \(\lambda=3.0\), linear \(\gamma_1\)}
                 & EF          & 0.09 & 14.2 & 9.7\\
\midrule
 & 2.29 & 17.6 & 7.9 & Vec<u32> & 0.33 & 20.0 & 8.6\\
Default &  &  &  &  CacheLineEF & 0.12 & 21.0 & 8.7\\
\multicolumn{4}{l}{\(\alpha=0.99\), \(\lambda=3.5\), cubic \(\gamma_3\)}
                           & EF & 0.09 & 21.2 & 9.6\\
\midrule
& 2.00 & 17.7 & 8.0 & Vec<u32> & 0.33 & 20.3 & 8.6\\
Compact&&& & CacheLineEF       & 0.12 & 20.9 & 8.6\\
\multicolumn{4}{l}{\(\alpha=0.99\), \(\lambda=4.0\), cubic \(\gamma_3\)}
                    & EF       & 0.09 & 21.7 & 9.7\\
\bottomrule
\end{tabular}
\vspace{-1em}
\end{table}

In \cref{tab:remap}, we compare the space usage and query throughput of the different remap
data structures for both the fast and compact parameters, for \(n=10^9\) keys. We observe that
the overhead of CacheLineEF is \(2.75\times\) smaller than a plain vector, and only \(40\%\) larger
than Elias-Fano encoding as implemented in the sux library \cite{sux-rs}.

The speed of non-minimal (PHF) queries that do not remap does not depend
on the remap structure used.

For \emph{minimal} (MPHF) queries with the for loop, with fast parameters, EF is significantly slower
(14.2~ns) with the fast parameters than the plain vector (12.5~ns), while
CacheLineEF (12.9~ns) is only slightly slower.
The difference is much smaller with the compact parameters, because the
additional computations for the cubic \(\gamma_3\) reduce the number of iterations
the processor can work ahead.
When streaming queries, for both parameter choices CacheLineEF is less than~0.1~ns slower than the
plain vector, while EF is 1~ns slower.

In the end, we choose CacheLineEF when using compact parameters, but prefer the
simpler and slightly faster plain vector for fast parameters.
Since $\alpha=0.99$ is close to $1$, the remap structure is not accessed much,
and the performance improvement of CacheLineEF over plain EliasFano coding is
not too large.

\begin{table}[t]
\caption{\label{tab:comparison}Performance comparison of MPHF methods on 300~million random
  string keys of uniform length between 10 and 50. Construction time is shown
  for 6 threads. A * indicates single-threaded timings (optimistic 6-fold
  speedup in parentheses). Near-optimal values in each column are bolded.%
  %TODO: UPDATE PtrHash, FMPH, and FMPHGO; check some others for consistency
}
\vspace{-0.5em}
\centering
\newcommand{\vname}[2]{\multirow{#1}{*}{\rotatebox[origin=c]{90}{#2}}}
\newcommand{\name}[2]{\multirow{#1}{*}{#2}}
\setlength{\tabcolsep}{5pt}
\scalebox{0.90}{
\begin{tabular}{lllrrr}
\toprule
&Approach & Configuration & \makecell[r]{\hspace{-2em}Space\\\hspace{-2em}bits/key} & \makecell[r]{Construction\\ 6t, ns/key} & \makecell[r]{Query\\ ns/query}\\
\midrule
\vname{5}{\small Bruteforce}
&\name{2}{SIMDRecSplit} & \(n{=}5\), \(b{=}5\) & 2.96 & \textbf{26} & 310\\[-0.1em]
&& \(n{=}8\), \(b{=}100\) & \textbf{1.81} & 66 & 258\\
\cmidrule{2-6}
&Bip. ShockHash-Flat & \(n{=}64\) & \textbf{1.62} & 2140* (357) & 201\\
\cmidrule{2-6}
&\name{2}{Consensus-RecSplit} & \(k=256\), \(\varepsilon=0.10\) & \textbf{1.58} & 521* (87) & 565\\[-0.1em]
&& \(k=512\), \(\varepsilon=0.03\) & \textbf{1.49} & 1199* (200) & 528\\
\midrule
\vname{6}{Fingerprinting}
&\name{2}{FMPH} & \(\gamma{=}2.0\) & 3.40 & 44 & 168\\[-0.1em]
&& \(\gamma{=}1.0\) & 2.80 & 69 & 236\\
\cmidrule{2-6}
&\name{2}{FMPHGO}& \(s{=}4\), \(b{=}16\), \(\gamma{=}2.0\) & 2.86 & 298 & 160\\[-0.1em]
&& \(s{=}4\), \(b{=}16\), \(\gamma{=}1.0\) & 2.21 & 423 & 212\\
\cmidrule{2-6}
&\name{2}{FiPS} & \(\gamma{=}2.0\) & 3.52 & 93* (\textbf{16}) & 109\\[-0.1em]
&& \(\gamma{=}1.5\) & 3.12 & 109* (\textbf{18}) & 124\\
\midrule
\vname{2}{\small Graph}
&\name{2}{SicHash} & \(p_1{=}0.21\), \(p_2{=}0.78\), \(\alpha{=}0.90\) & 2.41 & 48 & 149\\[-0.1em]
& & \(p_1{=}0.45\), \(p_2{=}0.31\), \(\alpha{=}0.97\) & 2.08 & 63 & 141\\
\midrule
\vname{16}{Bucket placement}
&CHD & \(\lambda{=}3.0\) & 2.27 & 1059* (177) & 542\\
\cmidrule{2-6}
&\name{4}{PTHash} & \(\lambda{=}4.0\), \(\alpha{=}0.99\), C-C & 3.19 & 403 & 77\\[-0.3em]
&& \ + HEM &  & 173 & \\
\cmidrule{3-6}
&& \(\lambda{=}5.0\), \(\alpha{=}0.99\), EF & 2.17 & 765 & 156\\[-0.3em]
&& \ + HEM  &  & 323 & \\
\cmidrule{2-6}
&\name{4}{PHOBIC} & \(\lambda{=}3.9\), \(\alpha{=}1.0\), IC-C & 4.14 & 62 & 116\\[-0.1em]
&& \(\lambda{=}4.5\), \(\alpha{=}1.0\), IC-R & 2.34 & 80 & 179\\[-0.1em]
%% && \(\lambda{=}6.5\), \(\alpha{=}1.0\), IC-R & \textbf{1.94} & 215 & 163\\[-0.1em]
&& \(\lambda{=}6.5\), \(\alpha{=}1.0\), IC-C & 2.44 & 220 & 108\\[-0.1em]
&& \(\lambda{=}7.0\), \(\alpha{=}1.0\), IC-R & \textbf{1.86} & 446 & 157\\
\cmidrule{2-6}
&\name{6}{\textbf{PtrHash}}
& \hspace{-2.85em}Fast\ \ \  \(\lambda{=}3.0\), \(\alpha{=}0.99\), $\gamma_1$, Vec
& 2.99 & \textbf{27} & \textbf{33}\\[-0.3em]
&&\ + streaming  &  &  & \textbf{16}\\
\cmidrule{3-6}
&& \hspace{-4.22em}Default\ \ \   \(\lambda{=}3.5\), \(\alpha{=}0.99\), $\gamma_3$, CLEF &
2.40 & \textbf{32} & \textbf{37}\\[-0.3em]
&&\ + streaming  &  &  & \textbf{23}\\
\cmidrule{3-6}
&& \hspace{-4.93em}Compact\ \ \  \(\lambda{=}4.0\), \(\alpha{=}0.99\), $\gamma_3$, CLEF &
2.12 & 63 & \textbf{35}\\[-0.3em]
&&\ + streaming &  &  & \textbf{23}\\
\bottomrule
\end{tabular}
}
\vspace{-1em}
\end{table}

\subsection{Comparison to Other Methods}
\label{sec:comparison}

In \cref{tab:comparison} we compare the performance of PtrHash against other methods on
short, random strings.
In particular, we compare against methods and configurations that are reasonably fast to construct:
SIMDRecSplit \cite{recsplit,recsplit-gpu},
Bipartite ShockHash-Flat \cite{shockhash,bipartite-shockhash},
Consensus-RecSplit \cite{consensus},
FMPH and FMPHGO \cite{fmph},
FiPS \cite{phf-thesis},
SicHash \cite{sichash},
CHD \cite{chd},
PTHash \cite{pthash,pthash-2},
and PHOBIC \cite{phobic}.
The specific parameters are based on Table 1 of \cite{phobic}, Table 8.1 of
\cite{phf-thesis}, and Table 3 of \cite{fmph}.
These results were obtained using the excellent MPHF-Experiments library
\cite{mphf-experiments} by Hans-Peter Lehmann. Construction is done on 6
threads in parallel when supported. By default, the framework queries
one key at a time. For PtrHash with streaming queries, we modified this to query
all keys at once.

\parag{Input.}
The input is 300~million random strings of random length between 10 and 50
characters. This input size is such that the MPHF data structures take around
75~MB, which is much larger than the 12~MB L3 cache.

\parag{PtrHash.} As expected, the space usage of PtrHash matches the numbers of \cref{tab:remap}.
In general, PtrHash can be slightly larger due to rounding in the number of
parts and slots per part, but for large inputs like here this effect is small.
Construction times per key are slightly slower than as predicted by
\cref{fig:construction}, while we might expect slightly faster construction due to the
lower number of keys. Likely, the slowdown is caused by hashing the input strings.
The hashing of input strings has a much worse effect on query throughput. In
\cref{tab:remap}, we obtained query throughput of 12~ns and 18~ns for the fast and compact
configurations when looping over integer keys, and as low as 8~ns when streaming queries. With
string inputs, these numbers increase to 33~ns resp. 35~ns when looping,
and 16~ns (resp. 23~ns) when streaming. A similar effect can be seen when comparing Tables 3
and 4 of \cite{fmph}.

\parag{Speed.}
We observe that PtrHash with fast parameters is the fastest to construct
alongside SIMDRecSplit (27~ns/key and 26~ns/key) and FiPS (16~ns/key, assuming optimal scaling to
6 threads),  resulting in around 3 bits/key for all three methods.
However, query throughput of PtrHash is \(9\times\) (SIMDRecSplit) resp.
\(3.3\times\) (FiPS) faster, going up to \(19\times\) resp.
\(6.8\times\) faster when streaming all queries at once.
Compared to the next-fastest method to query, PTHash-CC (HEM), PtrHash is twice
faster to query (or nearly \(5\times\) when streaming), is \(6.5\times\) faster to build, and
even slightly smaller.

With default parameters, PtrHash is \(2.1\times\) faster to query than the
fastest configuration of PTHash, and \(3.3\times\) faster when using streaming,
while being over \(5\times\) faster to construct.
Indeed, the speedup in query speed is explained by the fact that only a single
memory access is needed for most queries (compared to \(\geq 2\) for PtrHash-HEM
and PHOBIC), and generally by the fact that the code for querying is short.

\parag{Space.}
PtrHash with the fast parameters is larger (2.99 bits/key) than some other methods, but
compensates by being significantly faster to construct and/or query.
When space is of importance, the compact version can be used (2.12 bits/key).
This takes \(2.4\times\) longer to build at 63~ns/key, and has only slightly slower queries.
Compared to methods that are smaller,
PtrHash is over \(3\times\) faster to build than PHOBIC.
Consensus, SIMDRecSplit, and SicHash achieve smaller space of 1.58, 1.81, and 2.08~bits/key in
comparable time (63-87~ns/key), but again are at least \(3\times\) slower to query, or
over \(6\times\) compared to streaming queries.

\section{Conclusions and Future Work}
\label{sec:org9f032dd}
We have introduced PtrHash, a minimal perfect hash function that builds on
PTHash and PHOBIC. Its main novelty is the used of fixed-width 8-bit pilots that
simplify queries. To make this possible, we use \emph{hash-and-evict}, similar to
Cuckoo hashing: when there is no pilot that leads to a collision-free placement
of the corresponding keys, some other pilots are \emph{evicted} and have to search
for a new value.

The result is an MPHF with twice faster queries (37~ns/key) than any other method
(at least 77~ns/key) for datasets larger than L3 cache. Further,
due to its simplicity, queries can be processed in \emph{streaming} fashion, giving
another two times speedup (as low as 16~ns/key). At this point, the hashing of string inputs becomes a
bottleneck. For integer keys, such as $k$-mers, much higher throughput of up to
8~ns/key can be obtained, close to the 7.4~ns per cache line bandwidth, or
when using multiple cores even saturating the main memory~(2.5~ns/key).

\parag{Future work.}
A theoretical analysis of our method is currently missing. While
the hash-evict strategy works well in
practice, we currently have no relation between the bucket size \(\lambda\), load
factor \(\alpha\), and the number of evicts arising during construction.
Such an analysis could help to better understand the optimal bucket assignment
function, like PHOBIC \cite{phobic} did for the case without
eviction.
Possibly, the analysis of \cite[Section 5]{phobic-thesis} could
be extended to fully cover our method.

Second, the size of pilots could possibly be improved by further parameter
tuning. In particular we use 8-bit pilots, while slightly fewer or more
bits may lead to smaller data structures. An experiment with 4-bit pilots
was not promising, however.

To further improve the throughput, we suggest that more attention is
given to the exact input format. As already seen, hashing all queries at once
can provide significant performance gains via prefetching.  For string input
specifically, it is more efficient when the strings are consecutively packed in memory
rather than separately allocated, and it might be more efficient to explicitly
hash multiple strings in parallel.
More generally, applications should investigate whether they can be rewritten to take
advantage of streaming queries.
Furthermore, current throughput is limited by the fact that nearly every query
needs to fetch a new cache line. It would be interesting to design an MPHF that
only requires, say, half a cache line per query, or to disprove the
existance of such an MPHF.

Lastly, we refer the reader to PHast \cite{phast}, an MPHF that introduces a
number of interesting simplifications, leading to a datastructure is both
smaller and faster to query than PtrHash, although it is somewhat slower to construct.
It remains an open problem whether it is possible to construct an MPHF with
space within 0.1 bits/key from the lower bound that is as fast to query as
PtrHash and PHast.

\bibliographystyle{plainurl}
\bibliography{bibliography}

\newpage
\appendix
\crefalias{section}{appendix}

\section{Query Throughput}
\label{sec:throughput}

\subsection{Batching and Streaming}
\label{sec:orgabb5dd4}
\parag{Throughput.}
In practice in bioinformatics applications such as SSHash, we expect many
independent queries to the MPHF. This means that queries can be answered in
parallel, instead of one by one. Thus, we should optimize for query \emph{throughput}
rather than individual query latency. We report throughput as \emph{inverse
throughput} in amortized nanoseconds per query, rather than the usual queries
per second.

\parag{Out-of-order execution.}
An MPHF on \(10^9\) keys requires memory at least \(1.5\mathrm{\ bits}/\mathrm{key} \times 10^9
\mathrm{\ keys} = 188\)~MB, which is much larger than the L3 cache of size around
16~MB. Thus, most queries require reading a pilot from main memory (RAM), which usually
has a latency around 80~ns.
Nevertheless, existing MPHFs such as FCH \cite{fch} achieve an inverse throughput as
low as 35~ns/query on such a dataset \cite{pthash}.
This is achieved by \emph{pipelining} and the \emph{reorder buffer}.
For example, Intel Skylake CPUs can execute over 200 instructions ahead while waiting for memory
to become available \cite{measuring-rob,measuring-rob-skylake}. This allows the CPU to already start processing future
queries and fetch the required cache lines from RAM while waiting for the
current query. Thus, when each iteration requires less than 100 instructions
and there are no branch-misses, this effectively makes up to two reads in
parallel. A large part of speeding up queries is then to reduce the length of
each iteration so that out-of-order execution can fetch memory more iterations ahead.

\parag{Prefetching.}
Instead of relying on the CPU hardware to parallellize requests to memory, we can also
explicitly \emph{prefetch}\footnote{There are multiple types of prefetching
instructions that prefetch into a different level of the cache hierarchy. We
prefetch into all levels of cache using \texttt{prefetcht0}. Other prefetch
variants give similar results.} cache lines from our code.
Each prefetch requires a \emph{line fill buffer} to store the result before it is
copied into the L1 cache. Skylake has 12 line fill buffers~\cite{line-fill-buffer-skylake}, and hence can support up to 12 parallel
reads from memory.
In theory, this gives a maximal random memory throughput around $80/12 = 6.67$~ns per read
from memory, but in practice experiments show that the limit is 7.4~ns per read.
Thus, our goal is to achieve a query throughput of 7.4~ns.

We consider two models to implement prefetching: batching and streaming.

\parag{Batching.}
In this approach, the queries are split into batches (chunks) of size
\(B\), and are then processed one batch at a time.
In each batch, two passes are made over all keys.
In the first pass, each key is hashed, its
bucket it determined, and the cache line containing the corresponding pilot is prefetched.
In the second pass, the hashes are iterated again, and the corresponding slots are
computed.

\parag{Streaming.}
A drawback of batching is that at the start and end of each batch, the
memory bandwidth is not fully saturated.
Streaming fixes this by prefetching the cache line for the pilot \(B\) iterations
ahead of the current one, and is able to sustain the maximum possible number of
parallel prefetches throughout, apart from at the very start and end.

\clearpage
\subsection{Evaluation}
\label{throughput-evaluation}

\parag{A note on benchmarking throughput.}
To our knowledge, all recent papers on (minimal) perfect hashing measure query
speed by first creating a list of keys, and then querying all keys in the list,
as in \texttt{for key in keys \{ ptr\_hash.query(key); \}}. One might think this measures the average
latency of a query, but that is not the case, as the CPU will execute
instructions from adjacent iterations at the same time.
Indeed, as can be seen in \cref{tab:remap}, this loop can be as fast as 12~ns/key
for \(n=10^9\), which is over 6 times faster than the RAM latency of
around 80~ns (for an input of size 300~MB),
and thus, at least 6 iterations are being processed in parallel.

Hence, we argue that existing benchmarks measure (and optimize for)
throughput and that they assume that the list of keys to query is known in advance.
We make this assumption explicit by changing the API to benchmark all queries at
once, as in \texttt{ptr\_hash.query\_all(keys)}. This way, we can explicitly process
multiple queries in parallel.

We also argue that properly optimizing for throughput is relevant for
applications. SSHash, for example, queries all minimizers of a DNA sequence,
which can be done by first computing and storing those minimizers, followed by
querying them all at once.

We now explore the effect of the batch size and number of parallel threads on
query throughput.

\begin{figure}[t]
\centering
\includesvg[width=.95\linewidth]{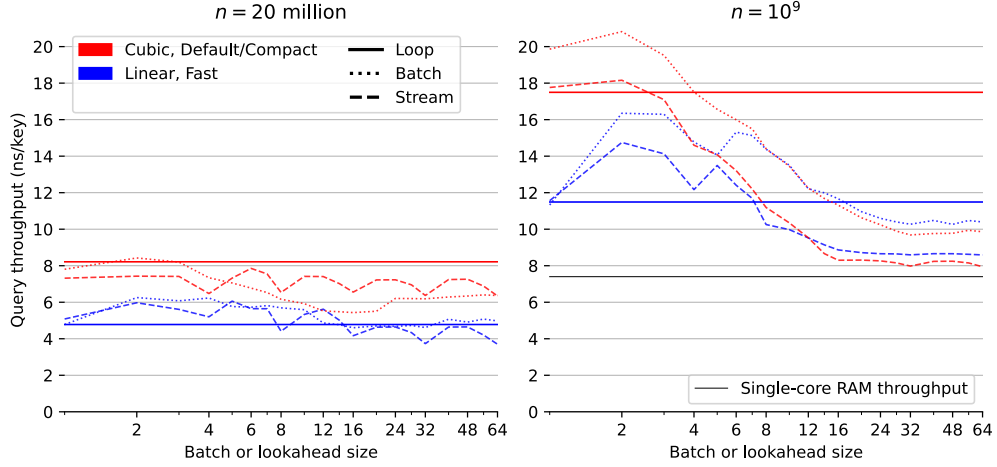}
\caption{\label{fig:batching}Query throughput of prefetching via batching
  (dotted) and streaming (dashed) with various batch/lookahead sizes, compared
  to a plain for loop (solid), for \(n=20\cdot 10^6\) (left) and \(n=10^9\)
  (right) keys. Blue shows the results for the fast parameters, and red for the
  compact parameters. Default parameters give performance nearly identical to
  the compact parameters, since the main differentiating factor is the use of
  $\gamma_1$ versus $\gamma_3$. All times are measured over a total of \(10^9\) queries, and for (non-minimal) perfect hashing only, \emph{without} remapping.}
\end{figure}

\parag{Batching and Streaming.}
In \cref{fig:batching}, we compare the query throughput of a simple for loop with the
batching and streaming variants with various batch/lookahead sizes. We see that
both for small \(n=20\cdot 10^6\) and large \(n=10^9\), the fast parameters yield
higher throughput than the compact parameters when using a for loop. This is
because of the overhead of computing \(\gamma_3(x)\). For small \(n\), batching and
streaming do not provide much benefit, indicating that memory latency is not a
bottleneck. However, for large \(n\), both batching and streaming improve over the
plain for loop. As expected, streaming is faster than batching here. For
streaming, throughput saturates when prefetching around 16 iterations ahead. At
this point, memory throughput is the bottleneck, and the difference between the
compact and fast parameters disappears. In fact, compact parameters with
\(\gamma_3\) are slightly \emph{faster}. This is because \(\gamma_3\) has a more skew
distribution of bucket sizes with more large buckets. When the pilots for these
large buckets are cached, they are more likely to be hit by subsequent queries,
and hence avoid some accesses to main memory.

For further experiments we choose streaming over batching, and use a lookahead
of 32 iterations.
The final throughput of 8~ns per query is very close to the optimal throughput of
7.4~ns per random memory read.

\label{sec:org8112f07}
\begin{figure}[t]
\centering
\includesvg[width=.95\linewidth]{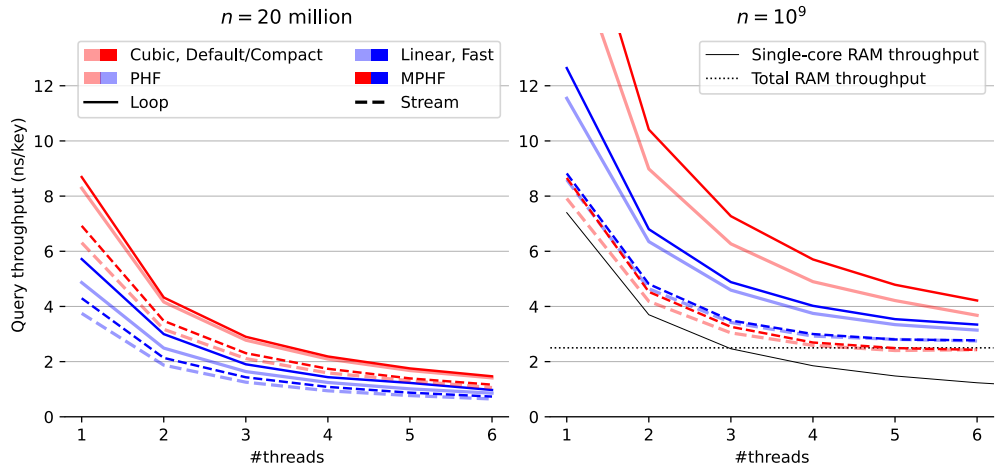}
\caption{\label{fig:throughput}In this plot we compare the throughput of a for loop
  (solid) versus streaming (dashed) for multiple threads, for both non-minimal
  (dimmed) and minimal (bright) perfect hashing. The left shows results for
  \(n=20\cdot 10^6\), and the right shows results for \(n=10^9\).
  In blue are the results for the fast parameters with $\gamma_1$, while results
  for the compact parameters with $\gamma_3$ are in red, which performs
  identical to the default parameters.
  On the right,
  the solid black line shows the maximum throughput based on 7.4~ns per random
  memory access per thread, and the solid black line shows the maximum
  throughput based on the total memory bandwidth of 25.6~GB/s.}
\end{figure}

\subsection{Multi-threaded Throughput.}
In \cref{fig:throughput} we compare the throughput of the fast and compact parameters for
multiple threads. When \(n=20\cdot 10^6\) is small and the entire data structure
fits in L3 cache, the scaling to multiple threads is nearly perfect. As
expected, minimal perfect hashing (bright) tends to be slightly slower than
perfect hashing (dimmed), but the difference is small. The fast \(\gamma_1\) is faster than
the compact \(\gamma_3\), and streaming provides only a small benefit over a for
loop.
For large \(n=10^9\), all methods converge towards the limit imposed by the full
RAM throughput of 25.6~GB/s. Streaming variants hit this starting at around 4
threads, and remain faster than the for loop. As before, the compact version is
slightly faster because of its more efficient use of the caches, and is even
slightly better than the maximum throughput of random reads to RAM.
Minimal perfect hashing is only slightly slower than perfect hashing.

\clearpage
\section{Sharding}
\label{sec:sharding}

When the number of keys is large, say over \(10^{10}\), their 64-bit (or 128-bit) hashes may not all fit
in memory at the same time, even though the final PtrHash data structure (the
list of pilots) would fit. Thus, we can not simply sort all hashes in
memory to partition them. Instead, we split the set of all \(n\) hashes into, say
\(s=\lceil n/2^{32}\rceil\) \emph{shards} of \(\approx 2^{32}\) elements each,
where the \(i\)'th shard corresponds to hash values in \(s_i:=[2^{64}\cdot i/s,
2^{64}\cdot (i+1)/s)\).
Then, shards are processed one at a time. The hashes in each shard are
sorted and split into parts, after which the parts are constructed as usual.
This way, the shards only play a role during construction, and the final
constructed data structure is independent of which sharding strategy was used.

\parag{In-memory sharding.}
The first approach to sharding is to iterate over the set of keys \(s\) times.
In the \(i\)'th iteration, all keys are hashed, and only those hashes in the
corresponding interval \(s_i\) are stored and processed.
This way, no disk space is needed for construction.

\parag{On-disk sharding.}
A drawback of the first approach is that keys are potentially hashed many times.
This can be avoided by writing hashes to disk. Specifically, we can create one
file per shard and append hashes to their corresponding file.
These files are then read and processed one by one.

\parag{Hybrid sharding.} A hybrid of the two approaches above only requires disk space
for \(D<s\) shards. This iterates and hashes the keys \(\lceil s/D\rceil\) times,
and in each iteration writes hashes for \(D\) shards to disk. Those are then
processed one by one as before.

\parag{On-disk PtrHash.}
When the number of keys is so large that even the pilots do not fit in memory, they
can also be stored to disk and read on-demand while querying. This is supported using $\varepsilon$-serde \cite{epserde,webgraph}.

\subsection{Evaluation}
\label{sec:sharding-eval}
We tested the in-memory and hybrid sharding by constructing PtrHash with default
parameters on \(5\cdot
10^{10}\) random integer keys on a laptop with only 64~GB of memory, using 6 cores
in parallel.
All 64-bit hashes would take 400~GB, so we use 24 shards of
around \(2^{31}\) keys, that each take 16~GB.
The final data structure takes 2.40 bits/key, or 15~GB in total, and the
peak memory usage is around 50~GB.

The in-memory strategy iterates through and hashes the integer keys 24 times, and takes
3098 seconds in total or 129~s per shard. Of this, 67~s (52\%) is spent on hashing
the keys, 14~s (11\%) is spent sorting hashes into buckets, and 45~s (35\%) is spent
searching for pilots.

The hybrid strategy is allowed to use up to 128~GB of disk space, and thus writes
hashes to disk in 3 batches of 8 shards at a time. This brings the total time
down to 2494~s (19\% faster), and uses 104~s per shard. Of this, an amortized 31~s (30\%) per shard is spent
writing hashes to disk, and 9~s (9\%) is spent reading hashes from disk, which
together is faster than the 67~s that was previously spent on hashing all keys.

\end{document}